# Remote control desk in Industry 4.0 for train driver: an ergonomics perspective

Emelyne Michel[a]*, Philippe Richard[a], Quentin Berdal[a]

*IRT Railenium - Railenium Research Institut, 180 rue Joseph-Louis Lagrange, Valenciennes 59540, France*

## Abstract

Remote control of trains will be an intermediary step before reaching full automation. In trains, use cases for remote control have been studied only for the past few years. This research presents a project about remote control for the next generation of trains in France and how we carry out the design of a new teleoperation desk for future remote train drivers. We present an Ergonomic Work Analysis used to precisely understand driver's activity. This analysis allowed us to identify the needs of future drivers and to propose ways to overcome one of the main problems that drivers will face when remotely driving a train: loss and degradation of sense. We explain how innovative technologies developed within the Industry 4.0 can offer solutions to problems faced with remote-control.

## 1. Introduction

A shift from manual control to automatic control under supervision, and ultimately to total autonomy has been noticed in the past years in technological development. High level of automation in the railway domain is already a reality. This is for example the case for the urban transport system in Grade of Automation 4 (GoA 4) in France (i.e., the first automatic subway in the world in Lille, and line 1 and 14 of the Paris subway) and all over the world. However, implementation of such system in the railway sector remains inaccessible in most cases in open areas because of challenges that still need to be addressed. Until these challenges are overcome and therefore fully autonomous train driving is reached, human operators will remain crucial for intermediate levels of automation. Meanwhile, growth of population and therefore increase of demand point out the necessity of expanding rail traffic and increase competitiveness in the respect of the low carbon national strategy in France for 2050. Furthermore, the Automatic Train Operator (ATO) system, a subsystem within the automatic subway train control, which performs automatic tasks such as brake and traction control, helps reduce financial investment. This reduction is allowed by the decrease in fixed infrastructure investment (e.g., building or opening new lines). Indeed, the ATO system would increase the capacity of railways by increasing the number of trains authorized to run on it.

* Corresponding author. Tel.: +33-782-201-819.
  E-mail address: emelyne.michel@railenium.eu

Remote control of trains or teleoperation seems to be a realistic solution. Teleoperation refers to the human capability to sense and manipulate a system from a distant location, in the context of this study, a train. Teleoperation in railways is a recent research topic and need for in-depht research in a panel of areas such as research dealing with human factors problematic raised with the introduction of a distance between the system and the human, technology and safety protocols. In this context, the TELLi project meets the requirement missing from the catalog of trains currently in circulation in France by conceiving a light, carbon-free train, and will be suitable both for the main railway lines and for what are called in France the "Lignes de Desserte Fine du Territoire" (LDFT, in English Low Density Lines) and which represent a third of the railway network (9 125 km in 2020). Within this project, the remote driving study is a work package of TELLi project, nicknamed TELLi-TC. TELLi-TC work package is part of a consortium composed of industrial and research institutes: SNCF, IRT Railenium and Ektacom. TELLi-TC meets the remote control needs identified for three use cases that had been identified and selected: technical movements (remote driving a train between a train station and a technical center with no passengers on board); technical movement within a technical center; remote control recovery autonomous GoA4 train (with or without passengers on board) after an accidental stopping on a rail line that does not longer allow ATO operation, for example, after an ATO failure.

Industry 4.0 refers to the fourth industrial revolution with a shift from mass production to customized production [1]. Industry 4.0 is a concept combining Internet of Things, Industrial Internet, Smart Manufacturing and Cloud based Manufacturing [2]. Thanks to all these prior technologies, remote control could counterbalance new problems that appear and can be seen as potential mitigation means to resolve them. Remote control fits a part of this fourth revolution as it gives an intermediary step before reaching full automation [3].

Teleoperation, which is now possible with the evolution of Innovative technologies that Industry 4.0 uses [4], consists in the ability to operate remotely through a set of techniques that transpose action capabilities (detection, manipulation, etc.) to a remote location in real time, thanks to sensory feedbacks picked up by sensors and transmitted to the teleoperator [5].

## 2. Perceptual sensation losses and degradation in teleoperation

Numerous researches show the importance of integrating teleoperation in parallel with automation development [6], [7], [8]. These researches illustrates the importance of sensors as an alternative solution to direct sensorial perception and the importance of identifying the type and formalism of information that will be communicated to remote drivers.

Since Chen et al. [9] review on teleoperation, several studies on this field emerged and highlighted challenges that teleoperators face [10], [11].

Driving a train is first a visual task: the driver must integrate different sources of visual information from external and internal sources [12]. While remotely driving a vehicle, particularly a train, the absence and alteration of sensory information causes an overexploitation of visual modalities compared to conventional driving (i.e. on-board driver) [13], [14]. To avoid this, it is primordial to value multimodal information sources on the driver desk design. Wickens' model [15] on Human information processing shows the utility to engage multimodal resources to facilitate task performing.

Multimodal interaction makes use of human senses in interacting with technology [16]. Human perceptual modalities (i.e., input) include visual, auditory, tactile, proprioceptive, vestibular, olfactory, and gustatory modalities. From these prior perceptual modalities follow several human output modalities (e.g., gaze, speech, sound, facial expressions, gesture of multiple body parts (head, hand, etc.), bio-electric measurements, exhalation, etc.). Perception is inherently multisensorial and cross-modal integration takes place between all senses [17]. Multiple researches [18], [19] show enhancement of user's satisfaction while other researches show a better task performance (faster time reaction and better accuracy).

Wickens' [15] multiple-resource model predicts the allocation of attentional resources when tasks are performed simultaneously, depending on their characteristics. This model focuses on mental workload and conflicts in information processing during task execution. Thus, for the same amount of information, tasks using simultaneously different sensory modalities (e.g., visual and auditory) will be better executed than tasks sharing a single sensory modality. Consequently, the use of a multimodal interface to convey information should lead to a better allocation of attentional resources, reducing mental workload. For example, transmitting visual information in parallel with a mainly visual task such as driving could cause an overload of the visual canal and consequently impact performance and security [20]. Using a multimodal interface to communicate information could enhance attentional resources repartition, and thus mental charge [19]. Therefore, communicating information with different sensorial modalities could prevent sensorial canal surcharge et enhance information processing capacity. Accordingly, situational awareness and performance is enhanced too [21], [18], [19].

Among these challenges, perceptual deprivation and interactions will be discussed in this paper. Among teleoperation challenges, deprivation and/or alteration of sensorial perception due to delocalization of the driver is part of the main investigation subject [22]. In this paper, we'll discuss how we considered these challenges into the design of a remote driver desk in our use cases.

*2.1. Visual system*

Visual modality is certainly the most important sense in human perception, indeed 70 % of the information processed by a human operator in a nominal situation is from visual modality to perceive shapes, colors, movement, depth and spatial relationships. It's the visual modality that is the most solicited when driving, other senses are supporting this predominant information. Compared to a traditional situation, remote situations generate an overexploitation of the visual modality which causes a security and a performance decrease [19].

Train drivers must monitor the driver desk and the external environment of the train simultaneously [23]. Remotely driving a train implies the use of an interface (camera) to visualize the surrounding environment. Depth, motion and speed perception, obstacle and sign detection are all examples of different aspects of the degradation of visual perception.

*2.2. Auditive system*

Beyond its primary role in communication, the auditory system alerts us to the proximity, location and direction of sound sources and informs us about our environment and the objects within it [24]. Assuming that the future remote drivers will be former on-board train drivers, sound transmission will be a necessity since it will allow them to rely on already established mental pattern [25].

In a remote driving task, system behavior can be perceived from single sources of information (e.g., system behavior could be diagnosed by sound alone) or combined sources of information (e.g., system behavior could be diagnosed by the combination of sound, visual scene, and inertial feeling). In remote driving, auditory signals will help improve situation awareness during a trip. Thus, the perception of the interaction between the train and its environment (e.g., braking by rail-wheel interaction a shock, acceleration, movement, slippage, etc.), the perception of sounds coming from inside the train, sounds related to weather conditions and sounds related to other environmental entities (e.g., the whistle of another train), are all indicators drivers could use to picture the situation they are in.

*2.3. Haptic system*

The haptic modality is the sense of touch, it includes kinesthetic feedback and tactile feedback [26]. Unlike auditory and visual modalities that provide precise temporal and spatial information, the transmission of haptic signals enables to reproduce the characteristics of object's surface [27].

The kinesthetic modality provides the individual with information about the position and movement of their body. Force feedback is a type of kinesthetic feedback often used in teleoperation [28], [29], [30], object manipulation [31], and driving [32].

The tactile modality includes electro tactile, thermal and vibrotactile feedbacks. The distinction between these feedbacks lies in the type of actuator used by the sensory system to provide information. Since thermal and electro tactile feedbacks can cause skin irritation, vibrotactile feedbacks are the most studied in literature. Vibrotactile signals can be used to navigate or to alert. In navigation, they allow to spatially guide individuals [33] or as a sensory substitution to provide directional cues, prevent hazards and provide alerts for accidents prevention [34]. In remote driving, it will be necessary to transmit this information with other haptic stimuli or with other senses, in order to support all the perceptions to which the haptic system contributes while driving (e.g., perception of movement, speed and acceleration or deceleration).

*2.4. Olfactory system*

The olfactory system provides nominal data of the different odors of the environment (chemical information). It is also an associative and emotional sense that triggers specific memories or reflexes contained in the smell memory (i.e., Proust effect). Although the olfactory system is less solicited while driving, it is still important to identify some specific anomalies whose detection could be delayed by teleoperation. For example, in remote driving, the driver could fail to detect a fire without a substitution of this modality.

*2.5. Vestibular system*

The vestibular system is in the inner ear and is an essential organ for an individual to perceive the position and displacements/orientation of their head [35]. This system consists of the semicircular canals and the otolith organs (i.e., the saccule and the utricle) which react only beyond a certain excitation threshold (respectively, the angular and linear perception thresholds) [36]. In driving, the sensory information provided by this system when combined with visual and somesthetic information allows an individual to orientate himself and perceive their own movement [37]. Since the remote driver will no longer be on board, visualizing the video of the environment can generate a sensorial conflict. Indeed, seeing the visual scene without any perception of movement normally captured by the inner ear could cause sensory conflicts with multiple symptoms (e.g., nausea) [38].

**3. Industry 4.0 and teleoperation**

In a simulation of the impact of Industry 4.0 on a typical automotive equipment plant, Berger [39] identified five key transformation levers for the industry [1]:

- The "virtual industrialization", which enables to virtually industrialize new product before their physical production.
- The "workflow automation", through the integration of autonomous vehicles or co-bots to perform tasks that are impossible to execute for humans due to the extent of the flow and packages, or parts, induced by the possibility to personalize the products.
- The "smart robots and machines" interconnect and self-correct without the input of operators for their operation. Operators must now supervise the process.
- The "predictive maintenance systems", which improves the planning of machine downtime due to their predictability, and consequently improve their utilization. The operators are part of the logic of diagnosis and problem solving.

- The "cyber-physical system and market", allowing for mass customization, the adaptation of the production plan in case of a variation in demand or a need for responsiveness.

Industry 4.0 relies heavily on the Internet of Object (IoT), which connects machines and systems to the internet, thus facilitating their remote control [40]. Sensors and IoT devices use wireless networks and sensors to collect real-time data on machine performance and production conditions, which can be analyzed and used for teleoperation [41], [42], [43]. In our use cases, a remote driver could, for example, monitor all the features of the train's safety systems, adjust its traction-braking from the driving desk. The data collected by IoT could help the driver to detect temperature variation, vibrations, or altitude changes [44], [45].

The analysis of data generated by IoT systems and carried out by Artificial Intelligence (AI) and Big Data could support a remote driver's decision-making. Sensory perceptions could be enriched by real-time recommendations or predictive analytics, improving their decision-making [46]. However, besides the fact that such technologies could reduce the influence of human factors, technologies that support automation systems can lead to inappropriate levels of trust [47], [48]. In the design of the remote driver's desk, we must ensure that an appropriate level of trust in the interface is reached (Trust calibration) [49]. Because of these drawbacks, the use of AI and Big Data is for now not considered in our project, but it is not excluded to add an AI-based system in the future with new advancements in terms of safety and in trust calibration research.

Mixed reality (MR) (augmented reality and virtual reality in particular) will increase operators' perceptual abilities. In a remote driving train, augmented information could be used to superimpose information of the environment video with speed limit or obstacle, and to identify signals identification on the video of the real environment [50], [51]. In addition, the immersion provided by mixed reality would improve the operators' sense of presence [52].

The more immersive and intuitive nature of human-machine interfaces will improve interactions between future remote drivers and interfaces. The use, for example, of touch interfaces, voice, or gesture commands could allow for more efficient monitoring and remote control of the train.

With these technological developments, these previous transformations (e.g., adding IA, MR, etc.) have an impact on operators, their activities, and the work organization [53]. In Industry 4.0, the need for low-skilled operators will decrease compared to what is required in traditional industry processes. Routine activities, once characterized by a low level of manual dexterity or social interaction requirements, will be replaced by technologies [54]. At the same time, Industry 4.0 will face an increase in the need for qualified operators who must demonstrate a high capacity to adapt, understand and master complex, interconnected and autonomous technical systems [53].

Another impact of the integration of Industry 4.0 principles can be seen in its effects on sensory perceptions with the modification of the interactions between operators and interfaces. In teleoperation, all sensory systems are affected by degradation, absence, or modifications in the sensory transmissions to the teleoperator. The design of the remote driver's desk intends therefore to consider these degradations and/or losses to prevent the decrease in performance that could be associated with them.

Romera et al. [55] suggested a typology of "augmented" human operators through different types of cooperation with intelligent systems (physical, sensory, cognitive) for physical or cognitive activities:
- A physically augmented operator 4.0 using co-bots and exoskeletons will see the performance of their manual operations increase and the risk of musculoskeletal disorders decrease.
- A sensory augmented 4.0 operator will be able to better perceive their environment and detect signals. Virtual or augmented reality visualization interfaces, for example, will allow for the increase of one's visual perception. Furthermore, receiving tactile signals to replace degraded or absent information can also increase the operator's workload.
- A cognitively augmented 4.0 operator will see their ability to process and interpret information increase, as well as their ability to solve problems. Diagnosis and analysis of the situation, as well as problem solving, will be improved by the contribution of 4.0 technologies [53].

Through these human enhancement, we can observe what are the different elements of Industry 4.0 that remote drivers will have access to, and which will be necessary to perform their teleoperation activities. An understanding of what the conventional train driving activity was necessary in order to transform this activity into a future remote-control activity with interface that will help the operator to enhance his performance. This understanding was achieved through an ergonomic analysis of the activity.

In the next section, we present the first results of the analysis of our understanding of the driving activity of a passenger train which helps us to identify needs for a future driving desk. Then, we discuss the future implementation of this activity's understanding. This paper focused on specific research question: What are the train drivers' needs in terms of perceptual information for their activity?

## 4. Materials and methods

The purpose of an Ergonomic Work Analysis (EWA) is to understand how an operator achieves the objectives set for him in each context, or to understand the reasons that prevented him from doing so. The dual activity regulation model [56] illustrates the dynamic interaction between internal and external determinants that will evolve according to the results of the activity. Activity is what is performed by the operator to perform a task. The activity covers observable behaviors or actions during the activity, but also falls within what is not directly observable such as mental activities [57]. Discourses about action realized by the operator are an observable form of activity, they accompanied the action and/or play a role in solving a problem.

The realization of an EWA enables us to understand the activity of an operator to ensure that the design of the remote driving desk meets the requirements of the future activity. As a first step, extractions of needs identified in literature analysis and prescription analysis of train drivers allowed for the EWA to focus on the activities such as brake and traction according to the understanding by the drivers of what is prescribed and the context in which they performed these tasks (i.e., external determinants of the situation). As a second step, for the EWA, just over 1:00 pm of train driving observations has been analyzed. The observations were made with a panel of nine French drivers (mean driving experience = 14.6 years; type of train driven: passenger or freight; type of routes: urban or rural) and in different weather conditions (rainy, sunny). All observation takes plan in a Regiolis workstation and lasted 1h27 on average. During this observation, train drivers were accompanied by their managers DPx (in french "Dirigeant de Proximité"). We had established an observation checklist, to obtain our subjective data, regarding perception of information about: search in the train's external environment, speed management, anomalies' identification, use of control-command, communication, HMI use, train localization. Because our goal was to identify required sensorial perceptions for train driving, we had to look for which information, when they used, and how to communicate them in the remote driver desk. Indeed, as we seen above, one of the main problems that drivers will face when remotely driving a train is loss and degradation of sense.

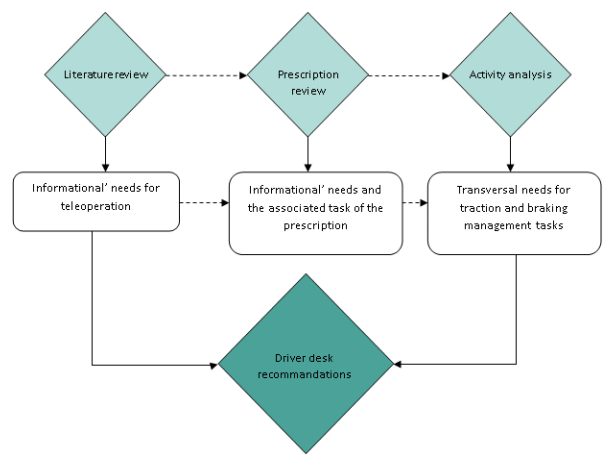

Fig. 1. The EWA methodology adopted for the study

## 5. Results

Data collected about the checklist observation allowed us, after a subjective analysis of the notes we took, to identify nine categories of transverse needs for many traction and braking management tasks performed by on-board drivers:

- Control and display management directly related to traction and braking: Traction and braking are the main activities for a driver. However, in remote control all driver sensations of speed or acceleration will be missing without the help of sensors that will send perceptual information to the drivers.
- Perception of the results of the driver's action: Remote drivers must be aware of the train's behavior and the implementation of their commands. The fact that a delay always exists between a command and its implementation by the train shows the importance for the remote driver to know what the train does.
- Situation awareness: The remote train driver must be aware of the situation and the environment around the train, especially since they are not any more physically there and lose the range of sensations they used to perceive.
- Sense of presence: Since the train driver is not physically in the train, the need to "be there" in the train is important for them to act and react as if they were in the train.
- Knowledge of the driving environment: the future remote driver will need to know the state of the train environment, the infrastructure, the presence of crossed trains. For example, it will be necessary for drivers to know when the train will meet a railway point.
- Train knowledge from the driver's desk: The driver needs to be informed of the state of the train, its localization and the restrictions applied on the railway area involved.
- Communications: Communication with anyone who usually interacts with the driver must be possible.
- Driver vigilance: Since the train driver is not physically in the train anymore the desk conception must support his vigilance and prevent vigilance degradation.
- Latency detection and management: Latency is inherent to teleoperation, but a remote driver must be aware of this latency and the design of the remote driving desk should enable the remote driver to adapt to it if this delay is stabilized as much as possible. Because un stabilized could prevent remote drivers to manage to adapt to delay.
- Human-machine cooperation: the new activity and future added sensors will require to design interfaces supporting the interaction between the train and the driver.

The perception of the results of the driver's action is a primary typology of needs. The perception of train speed at t time and the perception of speed evolution are examples of needs. The drivers interviewed reported the need to not systematically look at low abstraction level information but to related to sensational perception (e.g., technical displays) (translate from French "*but the driver will not spend his time looking at the speedometer, he does it according to the scrolling of the visual scene and sometimes he will look at the IV to check*", "*The reality is to rely on the directions of the scroll and the noise to have a feedback and while it has a feedback. It is the same as when you drive, you do not spend your time looking at the speedometer*.").

Train location is a must-have knowledge for train driving. Indeed, the drivers stressed the importance of always knowing the location of the train. In remote driving, the driver will have to know the location to decide which direction of travel to select, but also to know the restrictions that apply on the railway areas involved.

All these latter needs enable us to make recommendations to support the design of a remote driving desk. These recommendations focused on three hypotheses: H1: the visual modality is the main source of information and is supplemented by the auditory modality to transmit information to the remote driver; H1B: the visual modality remains the main source of information, but is also supplemented by the auditory and tactile modality to transmit information to the remote driver; H2: all modality can be used to transmit information to the remote driver (visual, auditory, tactile, vestibular, olfactory).

Furthermore, the specifications are divided into three steps of the driving activity in order to identify when the information is supposed to be presented to the teleoperator. The three steps of the driving activity are: Step 1, "Preparation of the mission, train, and departure" requires specifications concerning the needs related to the switch between the remote driver and the autonomous system; Step 2 "Driving" specifies recommendations concerning the perception needs of the effective speed at t time, acceleration, or deceleration; Step 3 "End of the mission" specifies recommendations related to train stop and end of mission, including desk lock-up and perception of train security.

## 6. Discussion

The first hypothesis implies that all the sensorial stimuli usually perceived by drivers are transformed into visual information. However, exclusive use of the visual modality would cause visual fatigue and could degrade the driver's performance.

The second hypothesis would undoubtedly limit the drawbacks observed with unimodal interfaces but would not completely avoid them. Multiple senses are involved in an operator's activity. These senses are not isolated, but are integrated into a single, coherent perception, thanks to the ability for humans to interact with multimodal sensorial stimulations. The third hypothesis therefore seems more in line with research on multimodality (e.g., avoiding single-channel overload, increasing information processing capacity, improving situation awareness, reducing mental workload, improving overall task performance) [21], [18], [19] than with the use of unimodal interfaces. However, as Meyer et al. [58] point out, the design of a multimodal system must ensure that the sensory modalities to be combined are carefully chosen, and that they are spatially, temporally and semantically congruent, and depending on the task to be performed [19], [59] individual differences [60] and the context of the task [61], [62], all of these in order to improve performance particularly in terms of speed and accuracy of detection. For example, the tactile modality could be used on the speed or traction-braking control-command to transmit information on the status of the commanded action. High force feedback when the commanded speed is far from the actual speed of the train, and low force feedback when it is close to the actual speed, would transmit to the remote driver information on the differential between their command and the actual speed. This information could be supplemented by a change in color of the speedometer according to the same rules. Another example of the use of multimodality would involve the use of auditory and visual modalities to transmit motion information. Auditory retransmission from the auditory sensors on the train, combined with the use of peripheral signals, would contribute to the remote driver's perception of movement.

In parallel with this study, a survey of sensor data that will be used to meet the identified needs has been carried out. This data should lead to an initial evaluation of the sensors to ensure that future remote drivers perceive and understand the information at an appropriate moment that will be transmitted to them.

Finally, Recommendations which had been made enabled us to design a first version of the driving desk. This first version will allow us to carry out a user test in order to evaluate, in a simplified remote driving activity, sensor's transmitted information and select sensory modalities. User test results will enable us to identify if former recommendations were adapted and if modifications should be established to enhance the driver desk.

The driving desk is for now designed (i.e., in terms of information and control command) to remotely drive a TLi with a kind of GoA2 level of automation. However, remote driving a train with higher level of automation will require to question the driving desk design and taking on new challenges such as authority transfers in a remote environment.

## 7. Conclusion and perspective

The TELLi-TC Work package contributes to a future deployment of train remote driving, independently of the deployment of autonomous train. The EWA provided an understanding of the driving activity currently carried out onboard trains by drivers. The use of a panel of sensory perceptions could be considered and integrated into the design of multimodal interfaces thanks to the technologies used in Industry 4.0. In addition, to enable future remote drivers to retrieve some of the information perceived onboard the train, innovative technologies could allow to go beyond what is perceived today by on-board drivers. Enhancements brought by innovative technologies, if they are properly designed, could tomorrow contribute to the improvement and support of the activity of future remote drivers.

## Acknowledgements

This research is part of the TELLi-TC Work package in a consortium of industrial partners: SNCF, Railenium and Ektacom. We thank them all for the opportunity to conduct this research and work on the development of a future remote driving desk in the railway. We would also like to show our gratitude to all train drivers who participated in this research.